\documentclass[10pt,final,twocolumn,journal]{IEEEtran}
\usepackage{url}
\usepackage{graphicx}
\usepackage[noadjust]{cite}
\usepackage{epstopdf}
\usepackage{cite}
\usepackage{caption}
\usepackage{subcaption}
\usepackage{authblk}
\usepackage{xcolor}
\usepackage{amsmath}
\usepackage{array}
\newcolumntype{P}[1]{>{\centering\arraybackslash}p{#1}}
\usepackage{enumitem}
\usepackage{booktabs}
\usepackage{adjustbox}
\usepackage[font=small,labelfont=bf]{caption}
\begin{document}

\title{Doping profile engineered triple heterojunction TFETs with 12 nm body thickness}

\author{ Chin-Yi Chen, Hsin-Ying Tseng, Hesameddin Ilatikhameneh, Tarek A. Ameen, Gerhard Klimeck,\\ Mark J. Rodwell, Michael Povolotskyi \vspace{-6.5ex}

\thanks{This was supported by National Science Foundation E2CDA Type I collaborative research on "A Fast 70mV Transistor Technology for Ultra-Low-Energy Computing" with the award number of 1639958 and semiconductor research corporation with its task ID of 2694.003. The use of nanoHUB.org computational resources operated by the Network for Computational Nanotechnology funded by the US National Science Foundation under Grant Nos. EEC-1227110, EEC-0228390, EEC-0634750, OCI-0438246, and OCI-0721680 is gratefully acknowledged. NEMO5 developments were critically supported by an NSF Peta-Apps award OCI-0749140 and by Intel Corp. This work also used the Extreme Science and Engineering Discovery Environment (XSEDE) at SDSC Dell Cluster with Intel Haswell Processors (Comet) through 50,000.0 SUs under charge number TG-ECS190009.}
\thanks{The authors are with the Department of Electrical and Computer Engineering, Purdue University, West Lafayette, IN, 47907 USA e-mail: r99941001@gmail.com}
}

\maketitle

\setlength{\textfloatsep}{12pt}
\setlength{\belowdisplayskip}{1.6pt} 
\setlength{\belowdisplayshortskip}{1.6pt}
\setlength{\abovedisplayskip}{1.6pt} 
\setlength{\abovedisplayshortskip}{1.6pt}
\setlength{\belowcaptionskip}{-12pt}
\vspace{-1.0\baselineskip}
\begin{abstract}
Triple heterojunction (THJ) TFETs have been proposed to resolve the low ON-current challenge of TFETs. However, the design space for THJ-TFETs is limited by fabrication challenges with respect to device dimensions and material interfaces. This work shows that the original THJ-TFET design with 12 nm body thickness has poor performance, because its sub-threshold swing is 50 mV/dec and the ON-current is only 6 $\mu A/\mu m$. To improve the performance, the doping profile of THJ-TFET is engineered to boost the resonant tunneling efficiency. The proposed THJ-TFET design shows a sub-threshold swing of 40 mV/dec over four orders of drain current and an ON-current of 325 $\mu A/\mu m$ with $V_{GS}$ = 0.3 V. Since THJ-TFETs have multiple quantum wells and material interfaces in the tunneling junction, quantum transport simulations in such devices are complicated. State-of-the-art mode-space quantum transport simulation, including the effect of thermalization and scattering, is employed in this work to optimize THJ-TFET design.
\end{abstract}

\section{introduction}
Power consumption in CPUs has impacted Moore's law significantly \cite{Bohr2011,Gonzalez1997}. An obvious solution to reduce the power supply is to replace the metal-oxide-semiconductor field-effect transistors (MOSFETs), which is limited by the Boltzmann tyranny, with new devices like the tunneling field-effect transistors (TFETs) \cite{Memisevic2016,Memisevic2017,Memisevic2018,Mohata2011,Sant2016,Appenzeller_2004,Appenzeller_2005,Ionescu_2011,Avci2015,Pang2019} and negative-capacitance field-effect transistors (NC-FETs) \cite{Khan_2011,Li2015}. However, these steep sub-threshold slope devices come with challenges that hinder their wide-spread applications. The primary challenge of TFETs is its low ON-current. TFETs are shown to suffer from low ON-current issue since the quantum tunneling probability is usually much lower than one \cite{Seabaugh2010}. 

The tunneling probability depends on several factors, such as tunneling distance, electric field, resonance conditions, and effective tunneling mass. Several approaches have been introduced to increase the tunneling probability based on optimizing these four factors. For example, in GaN-based heterojunction TFETs, the tunneling distance is reduced by engineering the band-diagram \cite{Tarek2019_alloy}; in a dielectric engineered TFET, the electric field at the tunneling junction is increased by using two different dielectrics \cite{Hesam_DETFET}; in a resonance-TFET, quantum resonances are used to increase the tunneling probability close to one \cite{Avci_2013}; in a Phosphorene-based TFET, the low effective tunneling mass increases the tunneling probability \cite{Ameen2016_BP}. 

A triple heterojunction (THJ-) TFET based on III-V materials allows an advantage in optimizing all of the factors mentioned above. A triple heterojunction reduces the tunneling distance using the  band diagram engineering. It also uses resonance tunneling to improve the tunneling probability in the ON-state and provides small effective tunneling mass due to III-V materials \cite{Huang2016_3HJ_TFET,Long2016_iciprm,Long2017,Huang2019}.

\begin{figure}[!t]
\center
\includegraphics[width=3.2in]{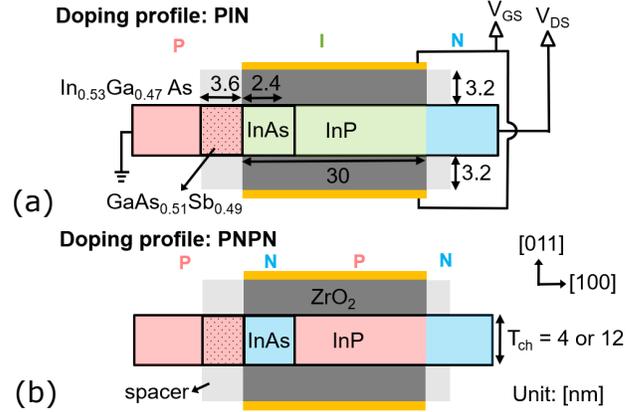}
\caption{Device design of a triple heterojunction TFET with (a) conventional doping profile (PIN) and (b) an optimized doping profile (PNPN).}
\label{schematic}
\end{figure}

Despite the benefits of THJ-TFETs, the fabrication constraints, such as device dimensions and material combinations, limit the performance of a THJ-TFET. For example, a 4~nm thick THJ-TFET with a conventional PIN doping profile shown in Fig.~\ref{schematic}(a) predicts an excellent performance; however, as the body thickness ($T_{ch}$) approaches a realistic value of 12~nm, the performance degrades. 

The reason for the high sensitivity of performance on body thickness in the original design is that, in PIN structures, the electric field (E) at the tunnel junction depends on both the depletion width ($W_{D}$) and the scaling length ($\lambda$), as shown in Eq. \ref{E_filed_eq} \cite{Sensitivity_2018,chinyi2018}: 

\begin{align}   
E \propto 1/ (W_{D}+\lambda(T_{ch})) 
\label{E_filed_eq}   
\end{align}
To address these issues in designing THJ-TFETs, the doping profile is engineered, as shown in Fig.~\ref{schematic}(b). By replacing the intrinsic part of the channel with the doped regions, the electric field is determined by the depletion width, which is not strongly dependent on $T_{ch}$ as shown in Eq. \ref{E_filed_eq2}.
\begin{align}   
E \propto 1/ (W_{D.source}+W_{D.channel}) 
\label{E_filed_eq2}   
\end{align}
Hence, the proposed design provides better performance for thicker devices by reducing the impact of thickness on the electrostatic profile.



The proposed design also considers fabrication technology constraints, including the limitation of the doping density in each material, the width of the strained quantum well, the crystal growth direction, and the channel material's choice to have a high-quality oxide interface. The proposed design shows the sub-threshold swing of 40 mV/dec over four orders of drain current. The high ON-current of 325 uA/um is achieved with a low supply voltage ($V_{DD}$) of 0.3 V.

The device design optimization is performed using the Nanoelectronics Modeling tool NEMO5~\cite{NEMO5_1,NEMO5_2}. The atomistic tight-binding method with a ten orbital $sp^3d^5s^*$ basis is used~\cite{Tan2016}. Carrier transport in THJ-TFETs is complex due to the presence of quantum wells in the tunneling region. The non-equilibrium  quantum mechanics of the system includes the electron-electron scattering and electron-phonon scattering of carriers in these quantum wells, tunneling process at multiple interfaces, and quantum confinement effects \cite{Luisier_2009,Lake_1997,Geng_2018,hsueh2016phonon,salfi2018valley,tankasala2018two,tankasala2015atomistic,Mazzola2020}. To capture these mechanisms, atomistic quantum transport simulation, including effective thermalization \cite{Huang2017, Ameen2017}, is necessary to evaluate the device performance. Since a real-space atomistic simulation is computationally challenging for devices with a large dimension, the atomistic mode-space approach developed in Ref.~\cite{chinyi_2020} is applied in this work. 

The paper is divided in four sections. The THJ-TFET device structure, that satisfies fabrication constraints, is presented in section II. The design principles for the THJ-TFET with the body thickness of 12~nm are discussed in section III. In section IV, we demonstrate the performance of the proposed THJ-TFET. And, the impact of the channel doping density is further discussed in section V.

\section{THJ-TFET device structure\\[-0.5em]}
Fig.~\ref{schematic} shows the double-gated ultra-thin-body (UTB) THJ-TFET studied in this work. Fig.~\ref{schematic}(a) is the THJ-TFET with a conventional PIN doping profile. It consists of a P-doped source, an intrinsic channel, and an N-doped drain. In the P-doped source, In$_{0.53}$Ga$_{0.47}$As and GaAs$_{0.51}$Sb$_{0.49}$ have the doping density of $N_{a}=5 \times 10^{19}$ cm$^{-3}$. InAs and InP channel are intrinsic. In the N-doped drain, InP has the doping density of $N_{d}=2 \times 10^{19}$ cm$^{-3}$. 

The UTB confinement direction is along $\langle$011$\rangle$ crystal direction and the electron transport direction is along $\langle$100$\rangle$ crystal direction. The electron transport direction is the same as the crystal growth direction to simulate the device structure fabricated by the vertical Fin-TFET technology \cite{Fujimatsu2013}. The choice of the materials in the heterojunction is compatible with current crystal growth technology limitations. The substrate is assumed to be InP such that InAs quantum well is under 3.41$\%$ bi-axial compressive strain, while In$_{0.53}$Ga$_{0.47}$As and GaAs$_{0.51}$Sb$_{0.49}$ are not strained since they are lattice-matched to InP substrate. The technology of growing In$_{0.53}$Ga$_{0.47}$As, GaAs$_{0.51}$Sb$_{0.49}$, and InAs on InP(100) substrate through molecular beam epitaxy (MBE) are all well-developed \cite{Tabata_1991,FENG2007121,lambert_1995}. The width of GaAs$_{0.51}$Sb$_{0.49}$ source quantum well is 3.6~nm. The width of strained InAs channel quantum well is 2.4~nm, which is less than the critical thickness of 4~nm \cite{Matthews_1974, Akazaki_1992}. The gate length is 30~nm, and the oxide thickness is 3.2~nm. The oxide material is assumed to be ZrO$_{2}$ with a relative dielectric constant of 15. The source is grounded. The drain is under the applied supply voltage ($V_{DD}$ = 0.3~V). The source to drain bias ($V_{DS}$) is 0.3~V. The spacer in this work is assumed to be air with the dielectric constant of 1.

Fig.~\ref{schematic}(b) is the proposed design with the same device structure as Fig.~\ref{schematic}(a) while the doping is changed to the PNPN doping profile. In the optimized PNPN doping profile, the InAs channel quantum well is doped to N-type with $N_{d}=2 \times 10^{19}$~cm$^{-3}$. The InP channel is doped to P-type with $N_{a}=2 \times 10^{19}$~cm$^{-3}$.

\section{Original THJ-TFET design principles}
\begin{figure}[!b]
\center
\includegraphics[width=2.8in]{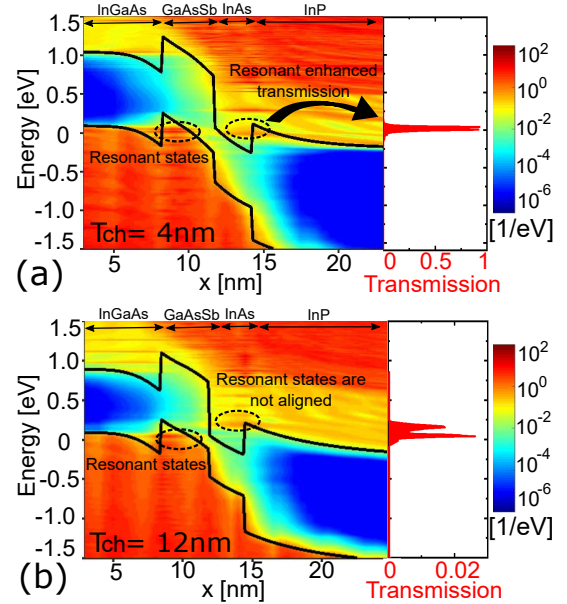}
\caption{Original THJ-TFET design principle: alignment of resonant states. (a) and (b) are the energy-resolved local density of states and transmission for 4 nm and 12 nm thick THJ-TFET when the device is operated in the ON-state.}
\label{LDOS_4nm}
\end{figure}
The design principle of the original THJ-TFETs is introduced in this section before describing the new design, PNPN-doped THJ-TFET. The energy-resolved local density of states (LDOS) and transmission probability for the original THJ-TFET with different body thicknesses ($T_{ch}$) are shown in Fig.~\ref{LDOS_4nm}. The doping profile is a conventional PIN doping profile shown in Fig.~\ref{schematic}(a). The LDOS in Fig.~\ref{LDOS_4nm} is calculated when the device is operated in the ON-state, where the gate to source bias ($V_{GS}$) is 0.3 V. In Fig. \ref{LDOS_4nm}(a), the alignment of the resonant states in the GaAsSb and InAs quantum wells results in the enhanced resonant tunneling such that the transmission probability is close to 1. On the other hand, in Fig.~\ref{LDOS_4nm}(b), when the body thickness increases to 12 nm, the resonant states are not aligned due to the worse gate control. The transmission, therefore, reduces 1$\sim$2 orders compared to the case of 4 nm body thickness. 

The key design rule of THJ-TFETs is to align the resonant states of two quantum wells in the tunneling junction and to introduce the resonant enhanced transmission. In the next section, the performance of 12 nm thick THJ-TFET is improved by aligning the resonant states through the proposed PNPN doping profile shown in Fig.~\ref{schematic}(b).


\section{Improved THJ-TFET with PNPN doping profile}
In this section, the performance of THJ-TFET with the PNPN doping profile is demonstrated. The PNPN doping profile was originally proposed for homojunction TFETs to improve electric field in the tunneling region \cite{Nagavarapu2008,Abdi2014,Baronia2017,abdi_2016}. It plays a more significant role in THJ-TFETs with thick body thickness. The PNPN doping profile can be engineered in THJ-TFETs not only to increase the electric field, but also to help aligning the resonant states that introduces the resonance tunneling.

\begin{figure}[!t]
\center
\includegraphics[width=2.6in]{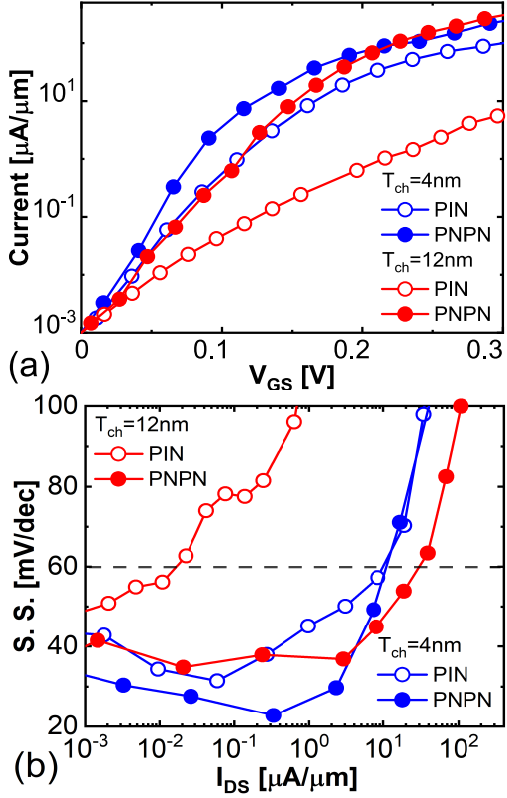}
\caption{(a) Transfer IV characteristics and (b) the S.S - $I_{DS}$ curve of a triple heterojunction TFET with a conventional doping profile (PIN) and the optimized doping profile (PNPN) for different body thicknesses ($T_{ch}$) of 4 nm and 12 nm.}
\label{4_12_nm_IV}
\end{figure}

Fig.~\ref{4_12_nm_IV}(a) compares the transfer characteristics of THJ-TFETs with 4 nm and 12 nm body thicknesses for different doping profiles. The gate to source bias ($V_{GS}$) are shifted to have a fixed OFF-current value of $10^{-3}$ $\mu A/\mu m$ at $V_{GS}$ = 0~V. For THJ-TFET with the PIN doping profile, when the body thickness increases from 4 nm to 12 nm, the loss of gate control dominates the performance such that the ON-current ($I_{ON}$) decreases by a factor of $\sim$16. 

\begin{table} [!b]
\begin{center}
\rule{0pt}{8pt}
\begin{tabular}{c c c c c c} 
 \hline
 \hline
 & &    \\ [-0.8em]
$T_{ch}$ &   & InGaAs  & GaAsSb & InAs & GaAsSb \\ 
\hline
 & &    \\ [-1em]
     & $E_{g}$ [eV]  & 0.9517 & 0.9871 & 0.7016 & 1.504 \\ [-0.6em]
4 nm & &\\ [-0.4em]
     & $\Delta E_{v}$ [eV]  & 0 & 0.4238 & 0.0456 & -0.3796 \\ 
\hline
 & & \\ [-1em]
     & $E_{g}$ [eV]  & 0.7993 & 0.8456 & 0.5010 & 1.3822 \\ [-0.6em]
12 nm & &\\ [-0.4em]
     & $\Delta E_{v}$ [eV]  & 0 & 0.4273 & 0.0670 & -0.3902 \\ 

 \hline
 \hline
\end{tabular}
\end{center}
\caption{The confined bandgap ($E_{g}$) and valence band off-set ($\Delta E_{V}$) of the heterojunction materials used in the design. The UTB's confinement direction is along $\langle$110$\rangle$. The atomistic tight-binding parameters used in this work are from \cite{Tan2016}.    }
\label{table:4_12nm_Eg}
\end{table}

However, for the THJ-TFET with the optimized PNPN doping profile, the same thickness increment is shown to improve the ON-current by $\sim$30$\%$. The reason is that, when the body thickness increases from 4 nm to 12 nm, the engineered built-in electric field in tunneling junction alleviates the effect of gate control degradation by a better doping profile design. At the same time, the decrease of the confined materials' bandgaps ($E_{g}$) enhances the ON-current \cite{chinyi2018}. The bandgaps of the materials used in the heterojunctions for different body thicknesses are listed in TABLE \ref{table:4_12nm_Eg}. The ON-current of the THJ-TFET with different body thicknesses and the doping profiles is summarized in TABLE \ref{table:4_12nm_Ion}. 

The sub-threshold swing (S.S.)-$I_{DS}$ curve for the THJ-TFETs with the PIN doping profile and the optimized PNPN doping profile for 4 nm and 12 nm thick THJ-TFETs are demonstrated in Fig.~\ref{4_12_nm_IV}(b). For a body thickness of 4~nm, the sub-threshold swing for the conventional PIN doping profile and the optimized PNPN doping profile does not show a significant difference. Both doping profiles exhibit decent performance. However, as the body thickness increases to 12~nm, the optimized PNPN doping profile retains its high performance, whereas the conventional PIN doping profile degrades drastically.

\begin{table} [!t]
\begin{center}
\rule{0pt}{8pt}
\begin{tabular}{ c c c c c} 
 \hline
 \hline
 & &    \\ [-0.5em]
 $T_{ch}$  & 4 nm  & 12 nm & 4 nm & 12nm \\ 
 & &    \\ [-1em]
 Doping profile  & (PIN) & (PIN) & (PNPN) & (PNPN) \\ 

\hline
 & & \\ [-1em]
 $I_{ON}$ [$\mu A/\mu m$] & 98  & 6 & 248 & 325              \\

 \hline
 \hline
\end{tabular}
\end{center}
\caption{The ON-current (I$_{ON}$) of the triple heterojunction TFET with body thicknesses of 4 nm and 12nm for both conventional PIN doping profile and the PNPN doping profile. The $I_{ON}$ is extracted at $V_{GS}$ = 0.3V.}
\label{table:4_12nm_Ion}
\end{table}

\begin{figure}[!t]
\center
\includegraphics[width=3.2in]{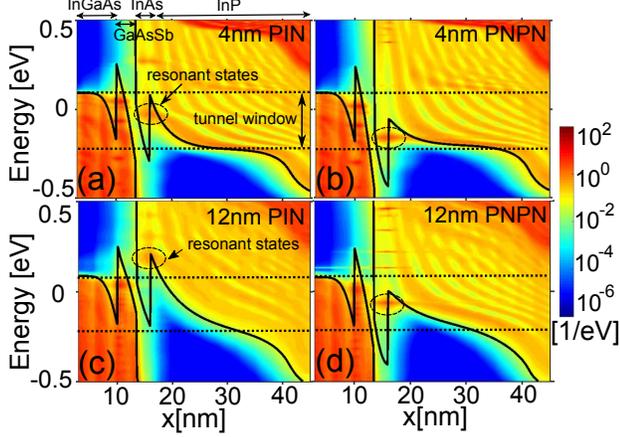}
\caption{Local density of states (LDOS) of the triple heterojunction TFETs with the body thickness/doping profile of (a) 4 nm / PIN, (b) 4 nm / PNPN, (c) 12 nm / PIN, and (d) 12 nm / PNPN. LDOS is calculated in the ON-state where $V_{GS}$ = 0.3 V.  }
\label{LDOS_4nm_12nm}
\end{figure}

To further understand why THJ-TFET with the optimized PNPN doping profile has a better performance comparing to the case of the traditional PIN doping profile, the local density of states (LDOS) at $V_{GS}$ = 0.3 V for different body thicknesses and different doping profiles are shown in Fig.~\ref{LDOS_4nm_12nm}.

When the channel thickness of the PIN-doped THJ-TFET increases from 4 nm to 12 nm, the quantum well states are misaligned, as shown in Fig.~\ref{LDOS_4nm_12nm}(a) and (c). The resonant states in the InAs channel quantum well are outside the tunneling window. The lack of resonance tunneling leads to significant degradation of the transmission probability and the ON-current. On the other hand, the optimized PNPN doping profile helps to retain the alignment of GaAsSb and InAs quantum well states in the 12 nm thick THJ-TFET, as shown in Fig.~\ref{LDOS_4nm_12nm}(d). The performance of 12 nm thick THJ-TFETs with the optimized PNPN doping profile is, therefore, similar to the case of a thinner channel thickness.

\begin{figure}[!b]
\center
\includegraphics[width=3.6in]{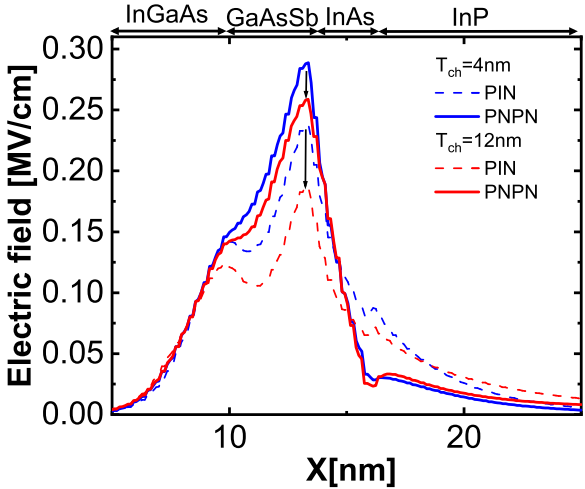}
\caption{Electric field along the channel for 4 nm and 12 nm thick triple heterojunction TFETs with both conventional PIN doping profile and optimzed PNPN doping profile. The electric field is obtained for the ON-state with $V_{GS}$ = 0.3~V.}
\label{e_field}
\end{figure}

The tunneling distance is determined by the electric field in TFET's tunneling region. Generally, the tunneling distance of a TFET with a conventional PIN doping profile is highly sensitive to the body thickness; a thinner device has a stronger gate control that leads to a smaller natural scaling length and hence a smaller tunneling distance~\cite{chinyi_2020,Ilatikhameneh2015,Hesam_FN}. Since the device with an optimized PNPN doping profile has no intrinsic region in the channel, the scaling lengths are dominated by the depletion width corresponding to the doping profile~\cite{Hesam_FN}. As a result, the optimized PNPN doping profile is not just engineered to increase the electric field in the tunneling junction; it also reduces the sensitivity of the performance to the body thickness. Fig.~\ref{e_field} shows the impact of doping profile and body thickness on the electric field along the channel. The peak electric field in PNPN doped THJ-TFETs has less dependence on the body thickness compared to the conventional PIN doped THJ-TFETs. 

\section{P-doped InP channel doping density}

\begin{figure}[!t]
\center
\includegraphics[width=2.7in]{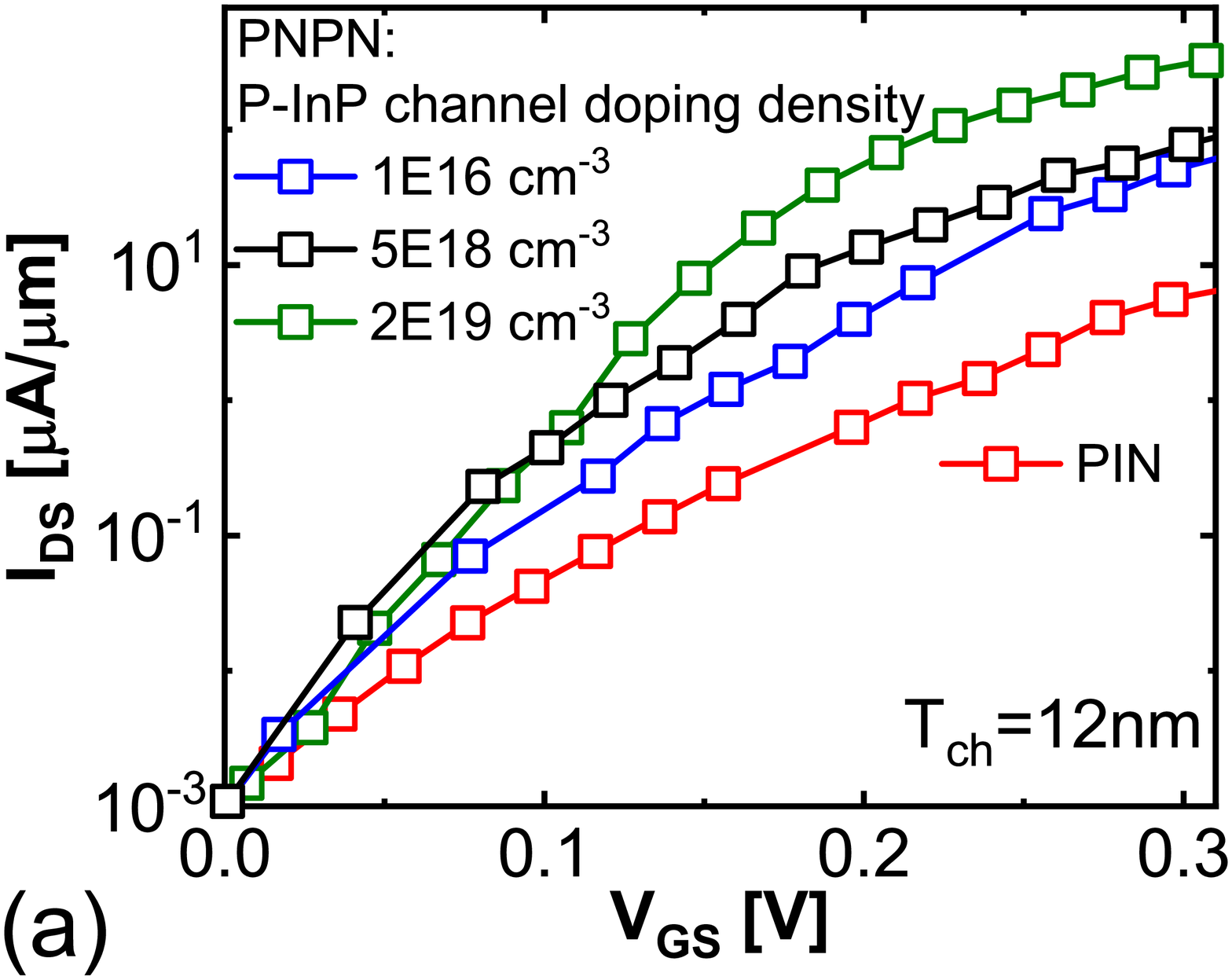}
\includegraphics[width=2.7in]{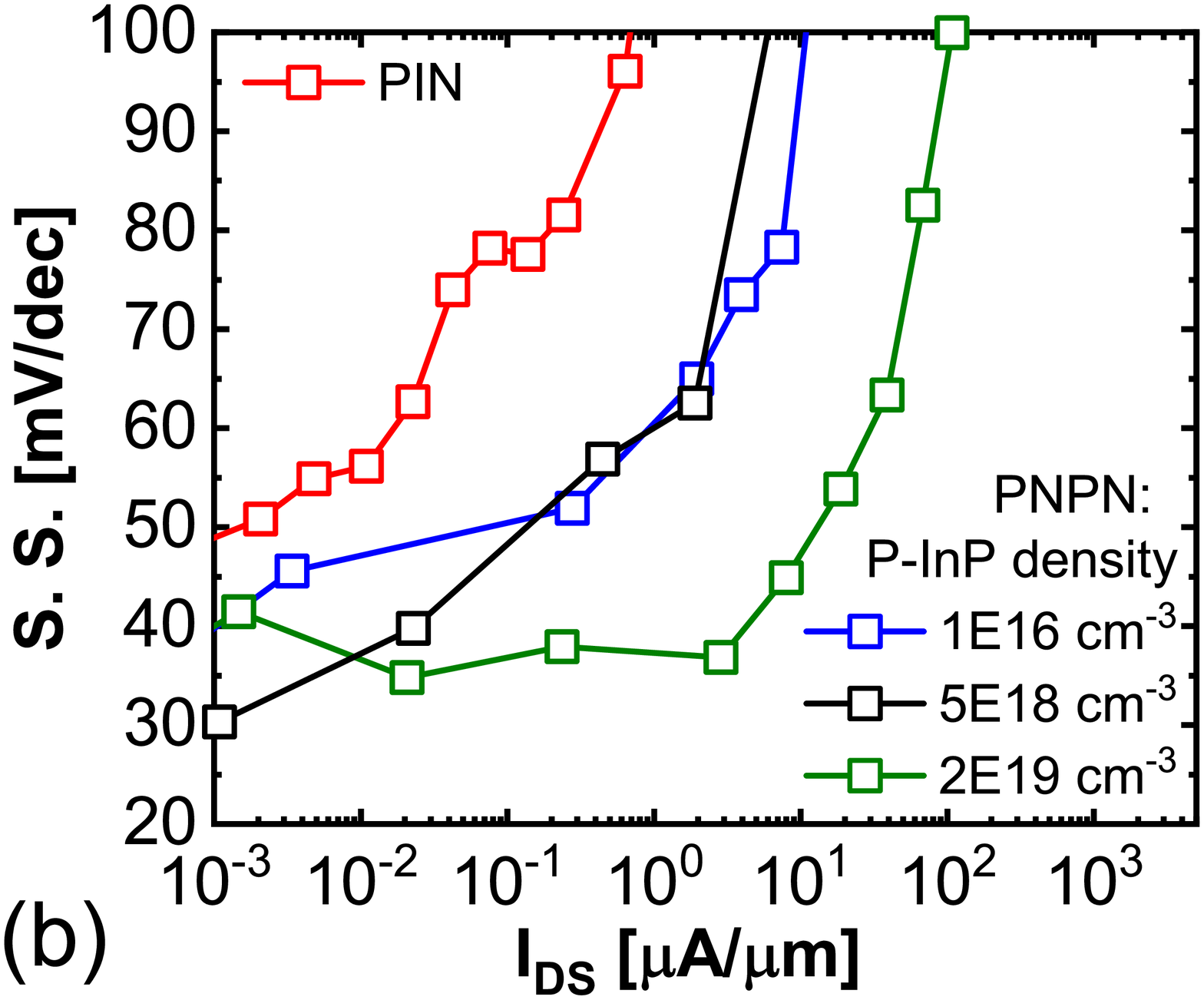}
\caption{(a) Transfer IV characteristics and (b) S.S. - $I_{DS}$ curve of the 12 nm thick triple heterojunction TFETs with the PIN and PNPN doping profile. The PNPN-doped TFETs with different P-InP channel doping densities are demonstrated. }
\label{compare_channel_doping}
\end{figure}

\begin{table} [!b]
\rule{0pt}{8pt}
\begin{tabular}{ c c c c c} 
 \hline
 \hline
 & &    \\ [-0.5em]
Doping profile  & PIN  & PNPN & PNPN & PNPN \\ 
 & &    \\ [-1em]
P-InP channel doping & -  & $1\times10^{16}$  & $5\times10^{18}$  & $2\times10^{19}$ \\ 
\hline
 & & \\ [-1em]
 $I_{ON}$ [$\mu A/\mu m$]  & 6  & 50 & 78 & 325              \\
 \hline
 \hline
\end{tabular}
\caption{I$_{ON}$ at $V_{GS}$ = 0.3 V of the 12 nm thick THJ-TFET with different doping profiles, such as conventional PIN doping profile and the PNPN doping profile, are listed.}
\label{table:12nm_ch_doping_Ion}
\end{table}

The benefit of having a P-N junction in the tunneling region is intuitive and is well-studied \cite{Nagavarapu2008, Abdi2014, Baronia2017, abdi_2016}. The electric field in the tunneling region is enhanced by P-N junction's built-in potential and therefore leads to a smaller tunneling distance and a larger transmission probability. The design rule of the P-N junction in the tunneling region is to maximize the doping density to achieve the maximum build-in potential. However, the role of the P-doped channel in the PNPN-doped THJ-TFET is not yet well understood. In this section, the doping density of the P-doped InP channel is studied. 

The transfer characteristics and S.S. - $I_{DS}$ curve of PNPN-doped THJ-TFET with different P-InP channel doping density are shown in Fig.~\ref{compare_channel_doping}. The body thickness of the device is 12 nm. The ON-current at $V_{GS}$ = 0.3~V is summarized in TABLE \ref{table:12nm_ch_doping_Ion}. The PNPN-doped THJ-TFET with P-InP channel doping density of $1\times10^{16}$ $cm^{-3}$ is the reference case to observe the improvement from applying the P-N junction in the tunneling region. The ON-current increases from 6 $\mu A/\mu m$ to 50 $\mu A/\mu m$ when the doping profile is replaced from PIN doping profile to PNPN doping profile with P-InP channel doping density of $1\times10^{16}$ $cm^{-3}$. 

Interestingly, the performance of 12 nm thick THJ-TFET improves slightly when P-InP channel doping density increases from $1\times10^{16}$ $\rm{cm}^{-3}$ to $5\times10^{18}$ $\rm{cm}^{-3}$. The ON-current increases from 50~$\mu \rm{A}/\mu \rm{m}$ to 78~$\mu A/\mu m$. The case with InP channel doped to $5\times10^{18}$ $\rm{cm}^{-3}$ shows sub-40 mV/dec S.S. for a limited range of drain current ($I_{DS}$). However, when the P-InP channel doping density further increases to $2\times10^{19}$ $\rm{cm}^{-3}$, the performance improves significantly. The ON-current of such case reaches 325~$\mu{\rm A}/\mu{\rm m}$. It exhibits the S.S. less than 40~mV/dec over four orders of magnitude in the drain current.

\begin{figure}[!t]
\center
\includegraphics[width=3.2in]{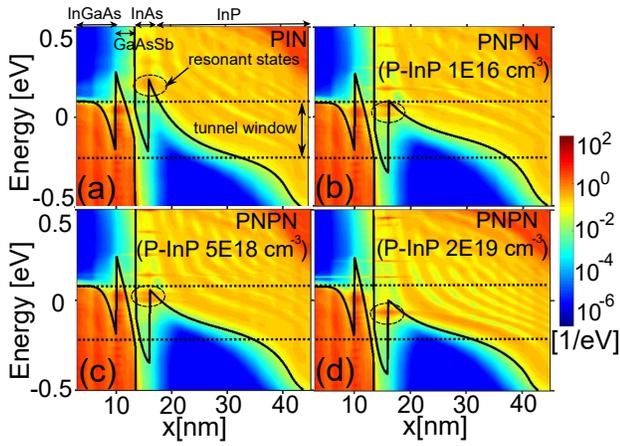}
\caption{Local density of states of 12 thick THJ-TFET with (a) the PIN doping profile and (b), (c), (d) the PNPN doping profile. The P-InP channel doping density in (b), (c), and (d) are $1\times10^{16}$, $5\times10^{18}$, and $2\times10^{19}$ $\rm{cm}^{-3}$, respectively. The local density of states is calculated for the ON-state with $V_{GS}$ = 0.3 V. }
\label{LDOS_diff_ch_doping}
\end{figure}

\begin{figure}[!t]
\center
\includegraphics[width=2.5in]{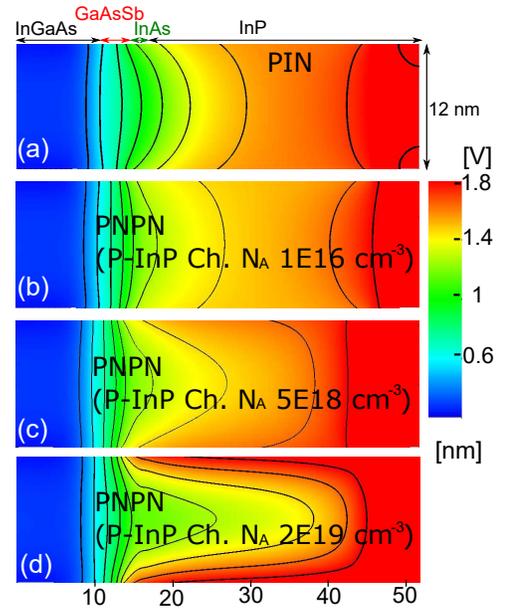}
\caption{Channel potential of the (a) PIN and (b), (c), (d) PNPN doping profile. The P-InP channel doping density in (b), (c), and (d) are $1\times10^{16}$ ${\rm cm}^{-3}$, $5\times10^{18}$ $\rm{cm}^{-3}$, and $2\times10^{19}$ $\rm{cm}^{-3}$, respectively. The potentials are obtained for the ON-state with $V_{GS}$ = 0.3 V. }
\label{2D_pot}
\end{figure}

\begin{figure}[!t]
\center
\includegraphics[width=3.5in]{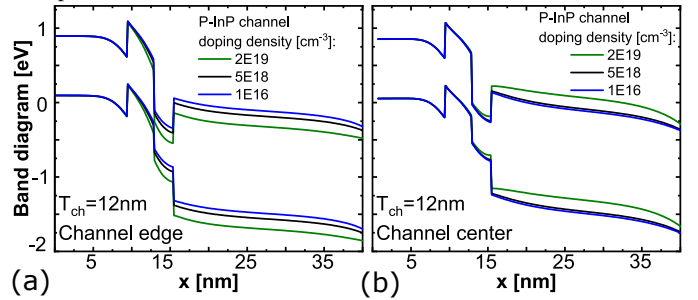}
\caption{The ON-state band diagram calculated at (a) the edge and (b) the center of the channel.}
\label{2D_pot_compare_edge_center}
\end{figure}

To further understand the impact of P-InP channel doping density, the ON-state local density of state (LDOS) is compared in Fig.~\ref{LDOS_diff_ch_doping}. In Fig.~\ref{LDOS_diff_ch_doping}, the LDOS, and the band-diagram are extracted at 1 nm away from the edge of the channel, where the potential is strongly affected by the gate bias. The resonant states in InAs quantum well are outside of the tunneling window in the case of PIN doping profile. On the other hand, for the cases of PNPN-doped THJ-TFET, the resonant states are all located inside the tunneling window regardless of different P-InP channel doping density. This indicates that the improvement when P-InP channel doped to $2\times10^{19}$ $\rm{cm}^{-3}$ comes from other factors other than the alignment of resonant states. The reason is illustrated through the 2D channel potential and the band diagram as shown in Fig.~\ref{2D_pot} and Fig.~\ref{2D_pot_compare_edge_center}. 


Fig.~\ref{2D_pot} shows the 2D channel potential for different channel doping profiles. In the case when P-InP channel doped to $2\times10^{19}$ $\rm{cm}^{-3}$, the 2D channel potential is significantly different from the cases with less channel doping density. The device with such high channel doping density is close to operated in the partially depleted regime; thus, a stronger vertical electric field toward the channel-oxide interface is observed. The strong vertical electric field lowers the channel barrier at the channel edge and increases the tunneling current.

 Fig.~\ref{2D_pot_compare_edge_center} (a) and (b) show the band diagram at the edge and the center of the channel, respectively. As the channel doping density increases to $2\times10^{19}$ $\rm{cm}^{-3}$, the valence band at the edge of the channel that close to the channel-oxide interface (depleted region) is further pushed down, which opens up a much lower resistant path comparing to other cases. Although the channel barrier at the center of the channel increases, it does not reduce the current, since the channel center is not the main path to conduct the current. As a result, a higher ON-current can be obtained as the channel doping density increases to $2\times10^{19}$ $\rm{cm}^{-3}$.

\section{Summary}
A triple heterojunction TFET design is proposed, considering fabrication constraints such as the channel thickness and the limitation in doping density of the materials. A triple heterojunction TFET with a conventional PIN doping profile is shown to degrade in performance when the body thickness increases from 4 nm to 12 nm. The new doping profile is engineered to increase the electric field in the tunneling junction and reduce the sensitivity of the performance to the body thickness. The ON-current of the optimized design reaches 325 $\mu {\rm A}/\mu\rm{m}$, and the S.S. is less than 40 mV/dec over four orders of magnitude in the drain current.

\bibliographystyle{ieeetr}
\bibliography{references.bib}

\end{document}